\documentclass{ws-ijmpa}
\usepackage[super,compress]{cite}
\usepackage{graphicx}
\begin{document}
\markboth{N. Abbasvandi, M. J. Soleimani, Shahidan Radiman, W.A.T. Wan Abdullah}{Quantum Gravity Effects On Charged Micro Black Holes Thermodynamics}

%
\catchline{}{}{}{}{}
%

\title{Quantum Gravity Effects On Charged Micro Black Holes Thermodynamics}

\author{Niloofar Abbasvandi}

\address{School of Applied Physics, FST, Universiti Kebangsaan Malaysia\\
Bangi, 43600, Malaysia\\
Niloofar@siswa.ukm.edu.my}

\author{M.J. Soleimani}

\address{National Center for Particle Physics, University of Malaya\\
KL, 50603, Malaysia\\
Department of Physics, University of Malaya, KL, 50603, Malaysia\\
Msoleima@cern.ch}

\author{Shahidan Radiman}

\address{School of Applied Physics, FST, Universiti Kebangsaan Malaysia\\
	Bangi, 43600, Malaysia\\
	Shahidan@ukm.edu.my}

\author{W.A.T. Wan Abdullah}

\address{National Center for Particle Physics, University of Malaya\\
	KL, 50603, Malaysia\\
	Department of Physics, University of Malaya, KL, 50603, Malaysia\\
	Wat@um.edu.my}

\maketitle

\begin{history}
\received{Day Month Year}
\revised{Day Month Year}
\end{history}

\begin{abstract}
The charged black hole thermodynamics is corrected in terms of the quantum gravity effects. Most of the quantum gravity theories support the idea that near the Planck scale, the standard Heisenberg uncertainty principle should be reformulated by the so-called Generalized Uncertainty Principle (GUP) which provides a perturbation framework to perform required modifications of the black hole quantities. In this paper, we consider the effects of the minimal length and maximal momentum as GUP type I and the minimal length, minimal momentum, and maximal momentum as GUP type II on thermodynamics of the charged TeV-scale black holes. We also generalized our study to the universe with the extra dimensions based on the ADD model. In this framework, the effect of the electrical charge on thermodynamics of the black hole and existence of the charged black hole remnants as a potential candidate for the dark matter particles are discussed.
\end{abstract}

\ccode{PACS numbers: 04.50.Gh, 04.60.-m}


\section{Introduction}
\label{}
The possibility of production of black holes \cite{argyres,bank,emparan} at particle colliders such as the Large Hadronic Collider (LHC) is one of the most exciting consequence of TeV-scale quantum gravity \cite{add}. Various theories of quantum gravity support the idea that near the Planck scale; the standard Heisenberg uncertainty principle should be reformulated by the so-called Generalized Uncertainty Principle (GUP)\cite{veneziano,Kempf,kempf01}. In particular, TeV-scale black hole physics \cite{meissner}, string theory, \cite{amati} and loop quantum gravity \cite{gary} indicate the existence of a minimum observable length and black hole Gedanken experiments support the idea in a fascinating manner \cite{scardigli,adler}. On the other hand, a test particle's momentum cannot be arbitrarily imprecise and there is an upper bound for momentum fluctuation based on the context of Doubly Special Relativity (DSR)\cite{ali,das}. Therefore, there is a maximal particle momentum. In this article, we study the effects of natural cutoffs encoded in GUPs on the thermodynamics of micro black holes with conserved electric charge during their formation and decay process \cite{calmet}. Firstly, we consider a GUP that admits just a minimal length and maximal momentum which we call GUPI. Secondly, we use a more general GUP that admits a minimal length, minimal momentum, and maximal momentum, which we call, GUPII. We study thermodynamics of charged TeV-scale black holes with extra dimensions in Arkani-hamed, Dimopoulos, and Dvali (ADD) model \cite{add} in the context of these GUPs. The corrections to charged micro black hole thermodynamic parameters such as Hawking temperature, Bekenstein-Hawking entropy, and radiation rate may have important consequences on TeV-scale black hole production at particle colliders. In this manner, we compare the obtained results by each of the above mentioned GUPs. Effects of extra dimensions, mass, and charge of the black holes is also discussed. The organization of this manuscript is as follows: in section 2, we introduce a generalized uncertainty principle with minimal length and maximal momentum, GUPI, and also GUPII which admits a minimal length, minimal momentum, and maximal momentum. In section 3, we obtain an expression for charged micro black holes Hawking temperature with GUPI and GUPII as well as the entropy and radiation rate corrected thermodynamic parameters based on the mentioned GUPs. In section 4, the electric charge effect on black hole thermodynamics and charged TeV-scale black hole remnant as a potential candidate for dark matter particle are discussed. The paper ended with conclusions in section 5.
\section{Generalized Uncertainty Principle}
\subsection{Minimal Length and Maximal Momentum (GUPI)}
\label{}
Most of the quantum gravity approach, indicate the existence of a minimal measurable length of the order of the Planck length, ${l_{pl}}\sim{10^{-35}}m$ \cite{maggiore,maggiore01}. The existence of a minimal measurable length modifies the Heisenberg uncertainty principle to the so-called Generalized Uncertainty Principle (GUP). The GUP framework is essentially restricted on the measurement precision of the particle's position, so that as the minimal position uncertainty could not be made arbitrarily small towards zero \cite{Kempf}. On the other hand, Doubly Special Relativity (DSR) theories \cite{camelia,amelino,amelino01} has considered that existence of a minimal measurable length would restrict a test particle's momentum to take any arbitrary values and therefore there exist an upper bound for momentum fluctuation \cite{magueijo,cortes}. It has been shown that there is a maximal particle's momentum due to fundamental structure of spacetime at the Planck scale. Based on this framework, the GUP that predicts both a minimal length and a maximal momentum can be written as follows \cite{ali,das}
\begin{equation} \label{1}
\Delta x\Delta p \ge \frac{\hbar }{2}\left( {1 - 2\alpha \langle p\rangle  + 4{\alpha ^2}\langle {p^2}\rangle } \right)
\end{equation}
or
\begin{equation} \label{2}
\Delta x\Delta p \ge \frac{\hbar }{2}\left( {1 - \alpha \langle \Delta p\rangle  + 2{\alpha ^2}{{\langle \Delta p\rangle }^2}} \right).
\end{equation}
The relation (1) and (2) can lead us to the following commutator relation:
\begin{equation} \label{3}
\left[ {x,p} \right] = i\hbar \left( {1 - \alpha p + 2{\alpha ^2}{p^2}} \right)
\end{equation}
where $\alpha$ is the GUP positive constant in the presence of both minimal length and maximal momentum. In the extra dimensional scenario based on the ADD model, the GUP can be written as follows \cite{nozari}:
\begin{equation} \label{4}
\Delta {x_i}\Delta {p_i} \ge \frac{\hbar }{2}\left( {1 - \alpha {L_{pl}}(\Delta {p_i}) + 2{\alpha ^2}L_{pl}^2{{(\Delta {p_i})}^2}} \right)
\end{equation}
where the Planck length in a model of the universe with large extra dimensions is defined as ${L_{pl}} = {(\frac{{\hbar {G_d}}}{{{c^3}}})^{\frac{1}{{d - 2}}}}$. Here, ${G_d}$ is the gravitational constant in $d$ dimensional spacetime, which is given by ${G_d} = {G_4}{L^{d - 4}}$ in the ADD scenario, where $L$ is the size of the extra dimensions. By saturating the inequality (4), a simple calculation gives,
\begin{equation} \label{5} 
\Delta {p_i} = (\frac{{\alpha {L_{pl}} + 2\Delta {x_i}}}{{4{\alpha ^2}L_{pl}^2}})\left( {1 \pm \sqrt {1 - \frac{{8{\alpha ^2}L_{pl}^2}}{{{{(\alpha {L_{pl}} + 2\Delta {x_i})}^2}}}} } \right)\
\end{equation}
So, the minimal position uncertainty has the value
\begin{equation} \label{6} 
\Delta {x_i} \ge \Delta {x_{\min }} = \alpha {L_{pl}}\left( {\frac{{2\sqrt 2  - 1}}{2}} \right).
\end{equation}
This is a new minimal observable length scale on the order of the Planck length.
\subsection{Minimal Length, Minimal Momentum, and Maximal Momentum (GUPII)}
In this section, we consider a more generalized uncertainty principle that admits a minimal length, a minimal momentum, and maximal momentum as well. As discussed in previous section, the minimal length comes from the finite resolution of spacetime points in the Planck scale, as a string cannot probe distances smaller than its length. In doubly special relativity theories, we consider the Planck energy (Planck momentum) as an additional invariant rather than the velocity of light. Therefore, the existence of a maximal momentum is in agreement with various theories of quantum gravity. The existence of a minimal momentum was developed by generalizing the Heisenberg commutation relation \cite{kempf02}. In this case, there is no rotation of a plane wave on a general curved spacetime \cite{hinrichsen,zarei}, for large distances where the curvature become important. In fact, the precision with which the corresponding momentum can be described and it can be expressed as a nonzero minimal uncertainty. For instance, one can obtain the harmonic oscillator energy spectrum in the GUP framework which implies maximal momentum uncertainty and minimal uncertainties in both position and momentum. It is known that quantum mechanical energy of its ground state, is nonzero and has a minimal value. In this case, the smallest uncertainty in momentum is not zero and can be considered nontrivially as the minimal momentum. Based on these arguments, as a consequence of small correction to the canonical commutation relations, one infers the following expression
\begin{equation} \label{7}
\Delta x\Delta p \ge \frac{\hbar }{2}\left[ {1 - 2\alpha (\Delta p) + 4{\alpha ^2}{{(\Delta p)}^2} + 4{\beta ^2}{{(\Delta x)}^2}} \right]
\end{equation} 
which in extra dimension can be written as follows
\begin{equation} \label{8}
\Delta {x_i}\Delta {p_i} \ge \frac{\hbar }{2}\left[ {1 - 2\alpha {L_{pl}}(\Delta {p_i}) + 4{\alpha ^2}L_{pl}^2{{(\Delta {p_i})}^2} + 4{\beta ^2}L_{pl}^2{{(\Delta {x_i})}^2}} \right]
\end{equation}
Here, $\alpha$ and $\beta$ are dimensionless, positive coefficients, and independent of $\Delta x$ and $\Delta p$ but in general they may depend on the expectation value of $\ x$ and $\ p$. The inequality (8) leads us to a nonzero minimal uncertainty in both position and momentum. It is easy to show  
\begin{equation} \label{9}
\Delta {x_i} \ge \Delta {x_{\min }} = \frac{{\hbar \alpha {L_{pl}}\left( {1 - 2\sqrt {1 - 12{\alpha ^2}{\beta ^2}L_{pl}^4} } \right)}}{{16{\alpha ^2}{\beta ^2}L_{pl}^4 - 1}}
\end{equation}
\begin{equation} \label{10}
\Delta {p_i} \ge \Delta {p_{\min }} = \frac{{\hbar \beta {L_{pl}}\left( {1 + 2\sqrt {1 - 12{\alpha ^2}{\beta ^2}L_{pl}^4} } \right)}}{{16{\alpha ^2}{\beta ^2}L_{pl}^4 - 1}}
\end{equation}
which these relations represent the existence of the minimal length and minimal momentum in presence of the extra dimensions based on the ADD model. Based on the generalized Heisenberg algebra, we suppose that operators of position and momentum obey the following commutation relation 
\begin{equation} \label{11}
\left[ {x,p} \right] = i\hbar \left( {1 - 2\alpha p + 4{\alpha ^2}{p^2} + 4{\beta ^2}{x^2}} \right).
\end{equation}
In this case, on the boundary of the allowed region, the curve is given by
\begin{equation} \label{12}
\Delta {p_i} = \left( {\frac{{\alpha {L_{pl}} + \Delta {x_i}}}{{4{\alpha ^2}L_{pl}^2}}} \right) \times \left( {1 \pm \sqrt {1-\frac{{(4{\alpha ^2}L_{pl}^2)(1 + 4{\beta ^2}L_{pl}^2{{(\Delta {x_i})}^2})}}{{{{(\alpha {L_{pl}} + \Delta {x_i})}^2}}}} } \right).
\end{equation}
In what follows, we use these two general forms of the GUP(I and II) as our primary input and construct a perturbational calculations to find thermodynamical properties of charged TeV-scale black hole and its quantum gravitational corrections. Here, we draw attention that since generalized Heisenberg uncertainty principle is a model independent concept \cite{hossenfelder}, the results which we obtain are consistent with any fundamental quantum gravity theory.
\section{Charged Micro Black Hole Thermodynamics}
\subsection{Hawking Temperature}
In order to characterize black hole, there are only three quantities namely, mass, $M$, electric charge, $Q$, and angular momentum, $J$ \cite{frolov,padmanaban}. In this manner, a charged black hole is the one which carries electric charge and the Schwarzschild solution is no longer valid. In this case, the Reissner-Nordstr\"{o}m geometry describes the empty space surrounding a charged black hole. On the other hand, the idea of the Large Extra Dimensions (LEDs) might allow studying interactions at Trans-Planckian energies in particle colliders such as LHC and the ADD model \cite{add} used new $d$ dimensional large space-like without curvature. Therefore, a natural candidate for TeV-scale charged black holes of higher dimensional is that of the Reissner-Nordstr\"{o}m $d$-dimensional solution of the Einstein field equation \cite{myers} given by
\begin{equation} \label{13}
d{s^2} = f(r){c^2}d{t^2} - {f^{ - 1}}(r)d{r^2} - {r^2}d{{\Omega}^{2} _{d - 2}} = {g_{\mu \nu }}d{x^\mu }d{x^\nu }
\end{equation}
where $\Omega _{d - 2}$ is the metric of the unit $S^{d - 2}$ as ${\Omega _{d - 2}} = \frac{{2{\pi ^{\frac{{(d - 1)}}{2}}}}}{{\Gamma (\frac{{d - 1}}{2})}}$, and
\begin{equation} \label{14}
f(r) = 1 - \frac{{{r_s}}}{{{r^{(d - 3)}}}} + \frac{{r_Q^2}}{{{r^{2(d - 3)}}}}.
\end{equation}
Here, the parameter ${r_s}$ is related to the mass $M$ of the black hole as 
\begin{equation} \label{15}
{r_s} = \frac{{8\pi {G_d}}}{{(d - 2){\Omega _{d - 2}}}}M
\end{equation}
where ${G_d} = {G_4}{L^{d - 4}}$ as mentioned in subsection 2.1 and the electric charge of the black hole is given by
\begin{equation} \label{16}
{Q^2} = \frac{{(d - 2)(d - 3)}}{{8\pi {G_d}}}r_Q^2
\end{equation}														
\begin{figure}[tpb]
\center
\includegraphics*[width=0.5\textwidth]{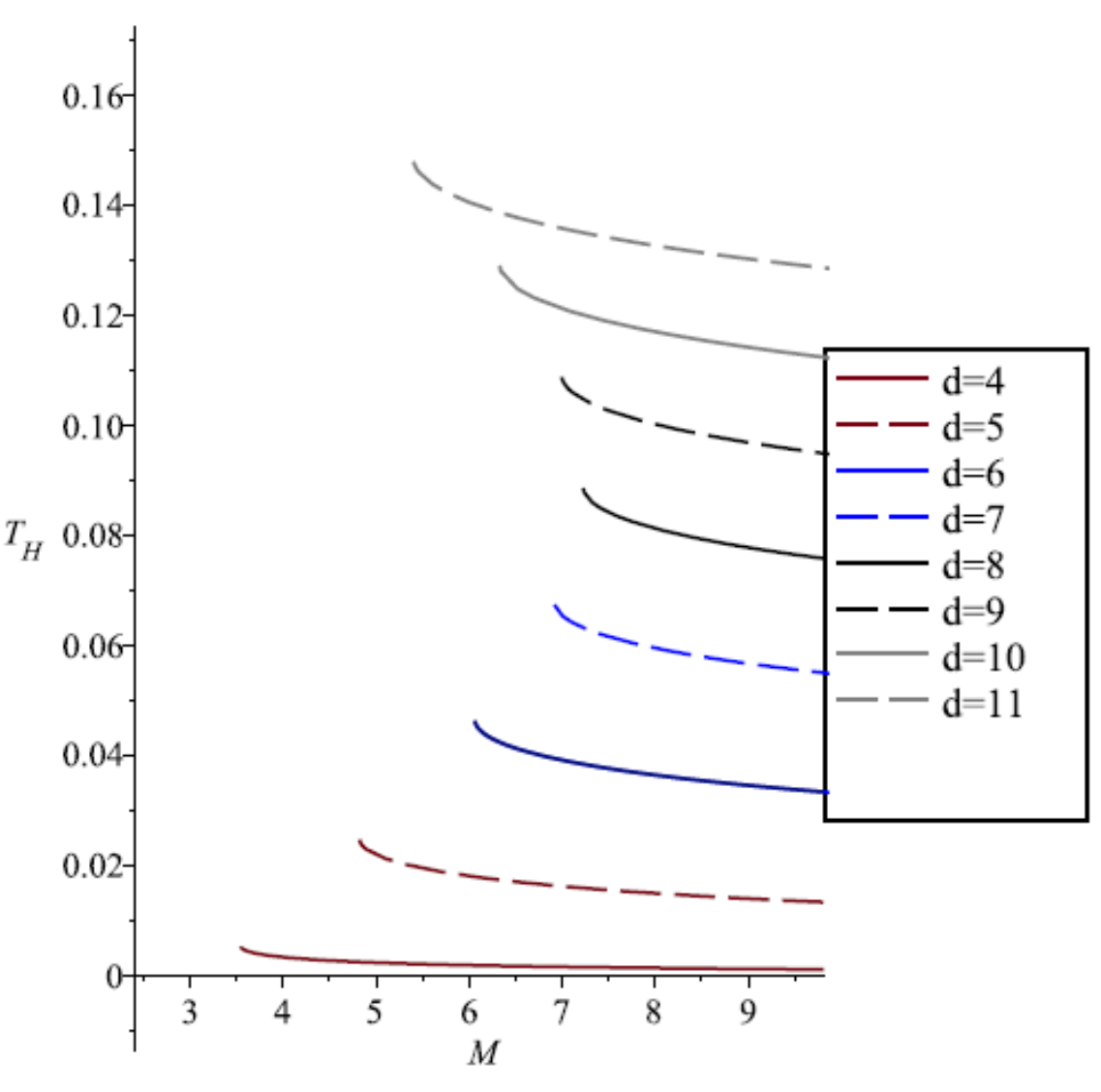}
\caption{Hawking Temperature with respect to the mass in terms of GUPI}
\label{fig1}
\end{figure}
where the outer horizon is situated at 
\begin{equation} \label{17}
r_h^{d - 3} = {r_s} + {\left( {r_s^2 - r_Q^2} \right)^{{\raise0.7ex\hbox{$1$} \!\mathord{\left/
				{\vphantom {1 2}}\right.\kern-\nulldelimiterspace}
			\!\lower0.7ex\hbox{$2$}}}}.
\end{equation}
In order to apply the original Bekenstein-Hawking formalism to the $d$-dimensional charged black holes, let us start with the first law of the black hole mechanics \cite{bekenstein,bardeen}, 
\begin{equation} \label{18}
dM = \frac{k}{{8\pi }}dA + \sum\limits_i^{} {{Y_i}} d{y_i}
\end{equation}
where $\sum\limits_i^{} {{Y_i}} d{y_i}$ are related to the work done on the black hole by an external agent. However, since Hawking radiation was discovered, it has been endowed with thermodynamic meaning, i.e. 
\begin{equation} \label{19}
dM = TdS + \sum\limits_i^{} {{Y_i}} d{y_i}.
\end{equation}
The first law is generalized to the electrically charged black holes as \cite{wald}
\begin{equation} \label{20}
dM = TdS + \mu dQ
\end{equation}
where $\mu $ plays the role of a chemical potential and $Q$ counts the number of charges. In general, the entropy of the black hole is assumed to be a function of its area, S=S(A) \cite{bekenstein}. Following the definition of thermodynamics, from (19) and (20), the temperature is expressed as 
\begin{equation} \label{21}
T = {\left( {\frac{{\partial M}}{{\partial S}}} \right)_Q} = \frac{{dA}}{{dS}} \times {\left( {\frac{{\partial M}}{{\partial A}}} \right)_Q} = \frac{{dA}}{{dS}} \times \frac{k}{{8\pi }}
\end{equation}
where the variable $Q$ is fixed. 
\\In this case, considering the black hole as $d$-dimensional cube, the position uncertainty should not be greater than a specific scale which is identified by twice radius of the horizon for a static spherically symmetric black hole such as Reissner-Nordström \cite{adler01}. Therefore, using GUPI leads to
\begin{equation} \label{22}
2{r_{h + }} \ge \Delta {x_i} \ge \frac{\hbar }{2}\left[ {\frac{1}{{\Delta {p_i}}} - \alpha {L_{pl}} + 2{\alpha ^2}L_{pl}^2(\Delta {p_i})} \right]
\end{equation}
which imposes constraint on the momentum uncertainty.
\\Following a heuristic argument \cite{adler01}, based on the usual Heisenberg uncertainty principle, uncertainty in the energy of the Hawking particles is $\Delta E \approx c\Delta p$. Here, we adopt the units $G = C = \hbar  = 1$. In this case, one deduces the following equation for the Hawking temperature of the black hole based on the LED scenario as
\begin{equation} \label{23}
{T_H} \approx \frac{{\left( {d - 3} \right)\Delta {p_i}}}{{4\pi }}
\end{equation} 
where, the constant, $\frac{{(d - 3)}}{{4\pi }}$, is a calibration factor in $d$-dimensional spacetime \cite{cavahli,cavahli01}. In this manner, the modified  Hawking temperature of the black hole based on the GUPI becomes 
\begin{equation} \label{24}
T_H^{GUPI} = \frac{{(d - 3)}}{{16\pi {\alpha ^2}{L^2}_{pl}}} \times \left[ {(4{r_{h + }} + \alpha {L_{pl}}) - {{({{(4{r_{h + }} + \alpha {L_{pl}})}^2} - 8{\alpha ^2}L_{pl}^2)}^{1/2}}} \right].
\end{equation}										
\begin{figure} [tpb]
\center
\includegraphics[width=0.5\textwidth]{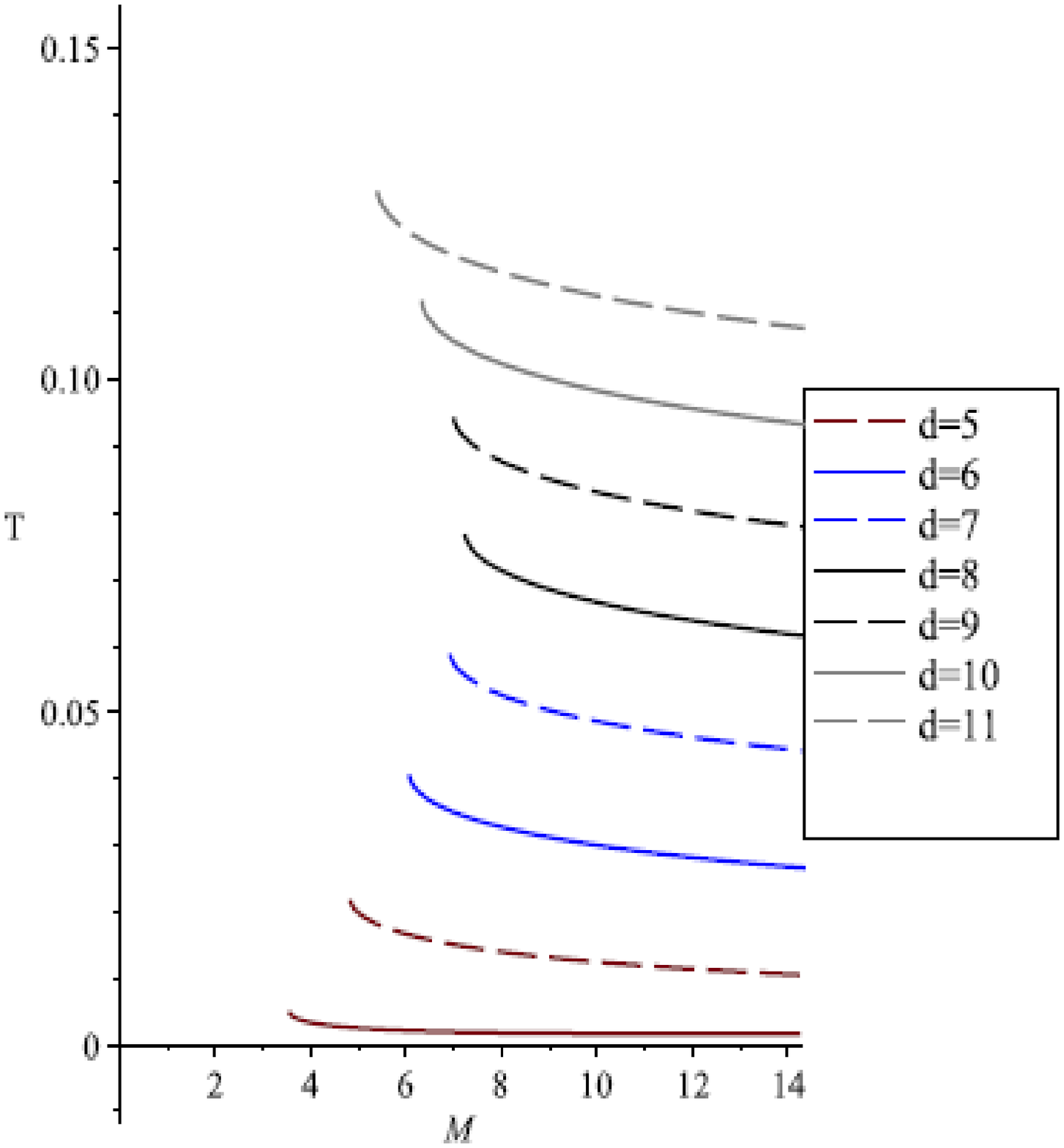}
\caption{Hawking Temperature with respect to the mass in terms of GUPII}
\label{fig2}
\end{figure}						
Based on the equation (24), GUPI gives rise to the existence of a minimal mass of the charged micro black hole given by
\begin{equation} \label{25} 
M_{\min }^{GUPI} = {\mkern 1mu} \frac{{\left( {d - 2} \right){\Omega _{d - 2}}}}{{16\pi {\mkern 1mu} {{\left[ {\left( {\frac{{2{\mkern 1mu} \sqrt 2  - 1}}{4}} \right)\alpha {L_{pl}}} \right]}^{d - 3}}}}\left( {{{\left[ {\left( {\frac{{2{\mkern 1mu} \sqrt 2  - 1}}{4}} \right)\alpha {L_{pl}}} \right]}^{2{\kern 1pt} d - 6}} + {\mkern 1mu} \frac{{8\pi {\mkern 1mu} {Q^2}}}{{\left( {d - 2} \right)\left( {d - 3} \right)}}} \right).
\end{equation}
The expression (25) shows that the Hawking temperature of the black hole  is only defined for $M \ge {M_{\min }}$.
In this case, the temperature of the black hole with minimum mass defined by (25) reaches a maximum value and reads
\begin{equation} \label{26}
\begin{array}{*{20}{l}}
{T_{\max }^{GUPI} = \frac{{(d - 3)}}{{4\pi {\alpha ^2}{L_{pl}}}}{M_p}^{\frac{1}{{(d - 3)}}}{{\left( {\frac{{8\pi M_{\min }^{GUPI}}}{{(d - 2){\Omega _{d - 2}}}}} \right)}^{\frac{1}{{d - 3}}}} \times {{\left( {1 + \sqrt {1 - \frac{{(d - 2)\Omega _{d - 2}^2{Q^2}}}{{(d - 3)8\pi M_{\min }^{GUPI}}}} } \right)}^{\frac{1}{{d - 3}}}}}\\
{\begin{array}{*{20}{c}}
	{}&{}
	\end{array} + \frac{{(d - 3)}}{{16\pi \alpha }}}
\end{array}
\end{equation}
In the standard picture of the micro black hole, the evaporation process can be divided into three characteristic stages \cite{gidding} as i)Balding phase which is the initial stage that black hole emits mainly gravitational radiations and sheds all the quantum numbers and multiple momenta apart from those determined by its mass, charge and angular momentum, ii)Evaporation phase which the black hole starts losing its angular momentum through the emission of the Hawking radiation, and iii)Planck phase as the final stage mainly the black hole mass approaches the true Planck scale as the black hole remnant with mass ${M_{min}}$. Within these stages, the charged TeV-scale black hole temperature increases through its evaporation process and the radius of the event horizon decrease in the framework of GUPI. This phase is also known as the Hawking phase. In the last stage, the temperature reaches to a finite temperature which is calculated by equation (26) (see Figure \ref{fig1} and Figure \ref{fig2}). However, as Figure \ref{fig3} shows, when electric charge, $Q$, increases, the minimum mass and its order of magnitude increase and the temperature peak displaces to the lower temperature. 
\\We also consider a more general uncertainty principle that admits a minimal length, a minimal momentum, and a maximal momentum, GUPII, to compute the corrected Hawking temperature of the charged black hole. Based on the equations (11), (12), and (23), in the same manner as GUPI, we obtain 
\begin{equation} \label{27}
T_H^{GUPII} = \frac{{(d - 3)(2{r_{h + }} + \alpha {L_{pl}})}}{{16\pi {\alpha ^2}L_{pl}^2}}\left[ {1 - \sqrt {1 - \frac{{4{\alpha ^2}L_{pl}^2(1 + 16L_{pl}^2{\beta ^2}r_{h + }^2)}}{{{{(2{r_{h + }} + \alpha {L_{pl}})}^2}}}} } \right].
\end{equation}
Therefore, the generalized uncertainty principle that admits a minimal length, a minimal momentum, and a maximal momentum, gives rise to the existence of a minimal mass of the charged black hole as
\begin{equation} \label{28} 
\begin{array}{l}
M_{\min }^{GUPII} = {\mkern 1mu} \frac{{\left( {d - 2} \right)\Omega }}{{16\pi {\mkern 1mu} }}\left( {{{\left( {{\mkern 1mu} \frac{{\left( {1 + 2{\mkern 1mu} \sqrt {1 - 12{\mkern 1mu} {\beta ^2}{\alpha ^2}L_{pl}^4} } \right){L_{pl}}\alpha }}{{32{\mkern 1mu} {\beta ^2}{\alpha ^2}L_{pl}^4 - 2}}} \right)}^{2{\kern 1pt} d - 6}} + {\mkern 1mu} \frac{{8\pi {\mkern 1mu} {Q^2}}}{{\left( {d - 2} \right)\left( {d - 3} \right)}}} \right)\\
\begin{array}{*{20}{c}}
{\begin{array}{*{20}{c}}
	{\begin{array}{*{20}{c}}
		{}&{}
		\end{array}}&{\begin{array}{*{20}{c}}
		{\begin{array}{*{20}{c}}
			{\begin{array}{*{20}{c}}
				{\begin{array}{*{20}{c}}
					{}&{}&{}
					\end{array}}&{}
				\end{array}}&{}
			\end{array}}&{}
		\end{array}}
	\end{array}}& \times 
\end{array}{\left( {{{\left( {\frac{{\left( {1 + 2{\mkern 1mu} \sqrt {1 - 12{\mkern 1mu} {\beta ^2}{\alpha ^2}L_{pl}^4} } \right){L_{pl}}\alpha }}{{32{\beta ^2}{\alpha ^2}L_{pl}^4 - 2}}} \right)}^{d - 3}}} \right)^{ - 1}}.
\end{array}
\end{equation}
\begin{figure} [tpb]
\center
\includegraphics*[width=0.5\textwidth]{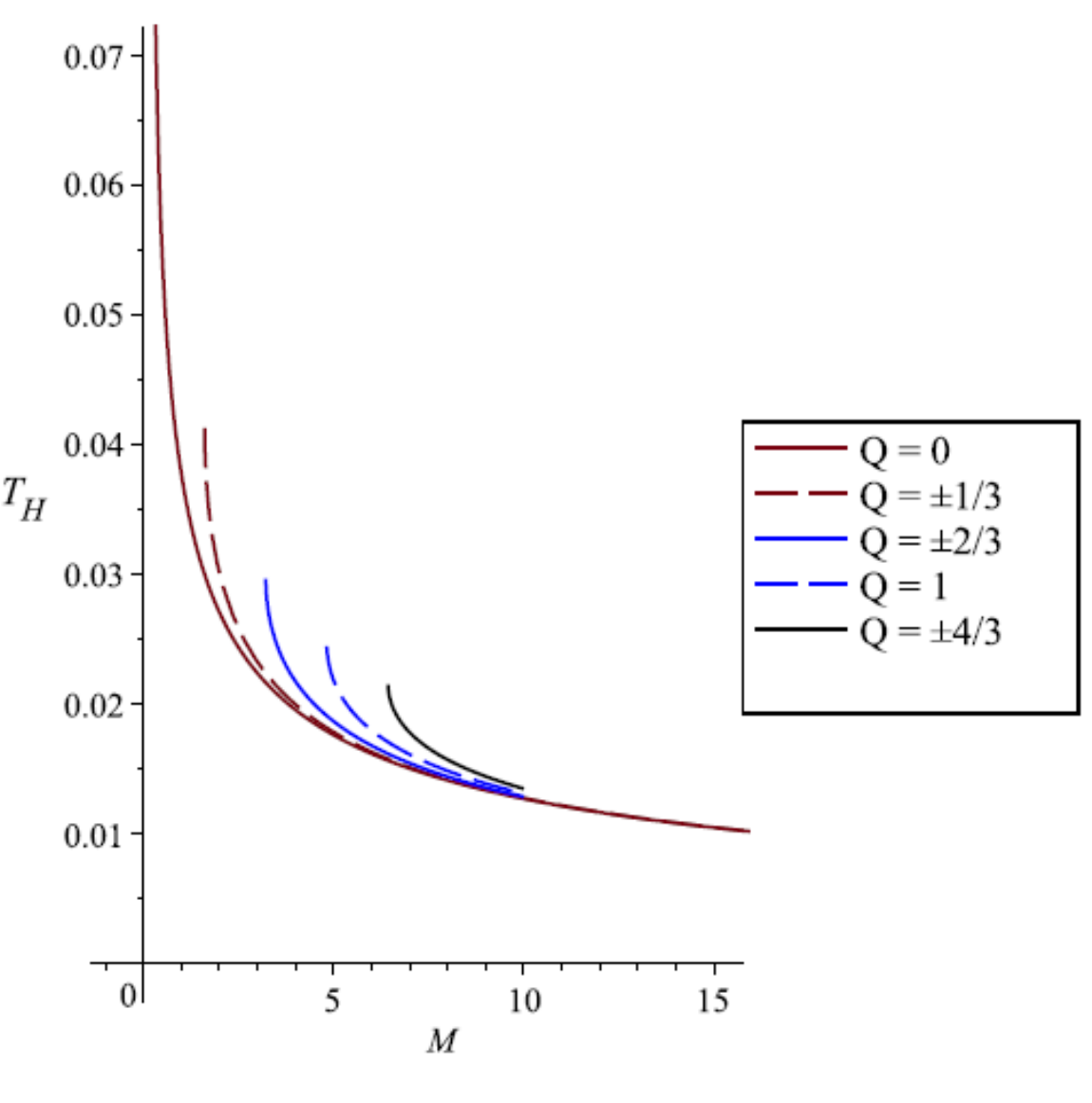}
\caption{Hawking Temperature for different amount of Charges}
\label{fig3}
\end{figure}		
Here, there is restriction on the range of the parameters $\alpha$ and $\beta$. The equations 27 and 28 shows that $\alpha$ and $\beta$ cannot take arbitrary value. Therefore, $\alpha$ and $\beta$ are related parameters which essentially depend on the aspects of the candidates for quantum gravity proposal.
The results show that in the large extra dimension scenario, the temperature of the charged black hole increases and leads to faster decay of the black hole. It is evident that in the large extra dimension scenario, the black hole remnant has mass more than its four dimensional counterpart. Therefore, in the generalized uncertainty principle framework, the quantum black holes are hotter, shorter lived and evaporate less than classical black holes.
\subsection{Entropy and Radiation}
It is well known that the Bekenstein-Hawking entropy is proportional to its horizon area which behaves in every way like a thermodynamic entropy \cite{hawking,bekenstein,vafa,carlip,solodukhin}. In order to find concrete form of the entropy of the charged micro black hole in presence of generalized uncertainty principle, we consider a particle captured by the black hole. Therefore, the loss of the information results in the increase of the entropy of the black hole. We obtain
\begin{equation} \label{29}
\Delta S \simeq \frac{{dS}}{{dA}}\Delta A
\end{equation} 
In this case, the inequality (4) can be rewritten in the Heisenberg uncertainty principle format, $\Delta {x_i}\Delta {p_i} \ge \hbar'$ where $\hbar'$ may be regarded as an effective Planck constant \cite{xiang}. Thus, the increase in area satisfies 
\begin{equation} \label{30}
\Delta A \ge \gamma \hbar'
\end{equation}
where $\gamma$ is a calibration factor.
\\The information of one bit is lost when a particle vanishes and the black hole specify increasing the entropy by ${(\Delta S)_{\min }} = \ln 2$. On the other hand, the lower bound of (29) gives the minimum increase in the horizon area. We then obtain 
\begin{equation} \label{31}
\frac{{dA}}{{dS}} \simeq \frac{{{{(\Delta A)}_{\min }}}}{{{{(\Delta S)}_{\min }}}} = \frac{{\gamma \hbar '}}{{\ln 2}}.
\end{equation}
By substituting (31) into equation (21) we get 
\begin{equation} \label{32}
T= \frac{k}{{8\pi }} \times \frac{{\gamma \hbar '}}{{\ln 2}}.
\end{equation}
In this manner, the standard result, $T = \frac{k}{{2\pi }}$, should be reproduced as $\alpha  \to 0$ which yields the calibration factor $\gamma  = 4\ln 2$. 
In this case, based on equation (31) the black hole entropy can be expressed as 
\begin{equation} \label{33}
S \simeq \int {\frac{{{{(\Delta S)}_{\min }}}}{{{{(\Delta A)}_{\min }}}}} dA = \frac{1}{4}\int {\frac{{dA}}{{\hbar '}}} 
\end{equation}
Since, we are dealing with the Reissner-Nordstr\"{o}m black hole of fixed charge, based on the equation (19), it is easy to show 
\begin{equation} \label{34}
S \simeq 2\pi \int {\frac{{dM}}{{k\hbar '}}}. 
\end{equation}
In order to obtain the modified entropy based on the GUPI, one should perform integration on $S$ which is physically reasonable to set ${M_{\min }}$ as lower limit of integration. Based on these arguments, and by substitution of equations (24) and (32) into (34), one can obtain 
\begin{equation} \label{35}
{S^{GUPI}} = \frac{{16\pi {\alpha ^2}L_{pl}^2}}{{d - 3}}\int_{{M_{\min }}}^M {dM \times {{\left[ {\left( {4{r_{h + }} + \alpha {L_{pl}}} \right)\left( {1 - \sqrt {1 - \frac{{8{\alpha ^2}L_{pl}^2}}{{{{(4{r_{h + }} + \alpha {L_{pl}})}^2}}}} } \right)} \right]}^{ - 1}}} 
\end{equation}
and similarly the entropy for the GUPII given 
\begin{equation} \label{36}
{S^{GUPII}} = \frac{{16\pi {\alpha ^2}L_{pl}^2}}{{d - 3}}\int_{{M_{\min }}}^M {dM \times } {\left[ {\left( {2{r_{h + }} + \alpha {L_{pl}}} \right)\left( {1 - \sqrt {1 - \frac{{4{\alpha ^2}L_{pl}^2(1 + 16L_{pl}^2{\beta ^2}r_{h + }^2)}}{{{{(2{r_{h + }} + \alpha {L_{pl}})}^2}}}} } \right)} \right]^{ - 1}}.
\end{equation}
The integral (36) can be solved numerically. One can use the semi classical entropy to measure the semi classical approximation validity. Therefore, the higher dimensional charged black hole remnants have less classical feature compared to their four dimensional counterpart. 
\\We now proceed to obtain the relation between emission rate of the charged TeV-scale black hole radiation and spacetime dimensions. It was shown \cite{emparan01} in $d$-dimensions, that the radiated energy by a black body of temperature $T$ and surface area $A$ is given by 
\begin{equation} \label{37}
\frac{{dE_d^{GUP}}}{{dt}} = {\sigma _d}AT_d^{GUP}
\end{equation}
which is based on the standard calculations of the statistical mechanics in higher dimensions, ${\sigma _d}$ is the $d$-dimensional Stefan-Boltzman constant, 
\begin{equation} \label{38}
{\sigma _d} = \frac{{{\Omega _{d - 3}}}}{{{{(2\pi )}^{d - 1}}(d - 2)}}\int_0^\infty  {\frac{{{Z^{d - 1}}}}{{{e^z} - 1}}} dz = \frac{{{\Omega _{d - 3}}}}{{{{(2\pi )}^{d - 1}}(d - 2)}}\Gamma (d)\zeta (d)
\end{equation}
with $\zeta (d)$ denoting the Riemann zeta function. Here, the outer event horizon is located at $r = {r_{h + }}$, and the area of the event horizon is ${A_d} = r_{h + }^{d - 2}{\Omega _{d - 2}}$. Therefore, using equations (24) and (27) as the modified Hawking temperature in the framework of the GUPI, and GUPII, we obtain
\begin{equation} \label{39} 
\frac{{dE_d^{GUPI}}}{{dt}} = \frac{{\left( {4{r_{h + }} + \alpha {L_{pl}}} \right){\Omega _{d - 3}}{\Omega _{d - 2}}(d - 3)}}{{16\pi {\alpha ^2}L_{pl}^2{{(2\pi )}^{d - 1}}(d - 2)}}\Gamma (d)\zeta (d)r_{h + }^{d - 2} \times \left( {1 - \sqrt {1 - \frac{{8{\alpha ^2}L_{pl}^2}}{{{{(4{r_{h + }} + \alpha {L_{pl}})}^2}}}} } \right)
\end{equation}
\\and
\begin{equation} \label{40}
\frac{{dE_d^{GUPII}}}{{dt}} = \frac{{\left( {2{r_{h + }} + \alpha {L_{pl}}} \right){\Omega _{d - 3}}{\Omega _{d - 2}}(d - 3)}}{{16\pi {\alpha ^2}L_{pl}^2{{(2\pi )}^{d - 1}}(d - 2)}}\Gamma (d)\zeta (d)r_{h + }^{d - 2} \times \left( {1 - \sqrt {1 - \frac{{4{\alpha ^2}L_{pl}^2(1 + 16{\beta ^2}L_{pl}^2r_{h + }^2}}{{{{(2{r_{h + }} + \alpha {L_{pl}})}^2}}}} } \right)
\end{equation}
\\These are complicated relation. Here, we highlight that the ${\sigma _d}$ changes very little with respect to the dimension. It was shown that most of the radiation goes into purely four dimensional fields, and the evaporation of the small black hole does not proceed as in a purely four dimensional theory \cite{emparan01}. In this case, the fact confirms that the rate of the energy which is radiated by black body with radius $R$ and temperature $T$ is roughly independent of the dimensions even though higher dimensional spacetime have infinitely many more modes due to the excitations in the extra dimensions. Here, we consider the case of $d = 4$, $d = 7$, and $d = 10$. Since ${\sigma _4} \simeq 0.082$, ${\sigma _7} \simeq 0.062$ and ${\sigma _{10}} \simeq 0.097$, one can compute numerically the following ratios, 
\begin{equation} \label{41}
\frac{{(\frac{{dE_4^{GUPI}}}{{dt}})}}{{(\frac{{dE_7^{GUPI}}}{{dt}})}} \simeq 9.89,\frac{{(\frac{{dE_4^{GUPII}}}{{dt}})}}{{(\frac{{dE_7^{GUPII}}}{{dt}})}} \simeq 9.89,\frac{{(\frac{{dE_4^{GUPI}}}{{dt}})}}{{(\frac{{dE_{10}^{GUPI}}}{{dt}})}} \simeq 10.84,\frac{{(\frac{{dE_4^{GUPII}}}{{dt}})}}{{(\frac{{dE_{10}^{GUPII}}}{{dt}})}} \simeq 10.84
\end{equation}
and
\begin{equation} \label{42}
{\left.{\frac{{(\frac{{dE_4^{GUPI}}}{{dt}})}}{{(\frac{{dE_7^{GUPI}}}{{dt}})}}} \right|_{Q = 0}} \simeq 8.76,{\left.{\frac{{(\frac{{dE_4^{GUPII}}}{{dt}})}}{{(\frac{{dE_7^{GUPII}}}{{dt}})}}} \right|_{Q = 0}} \simeq 8.76,{\left. {\frac{{(\frac{{dE_4^{GUPI}}}{{dt}})}}{{(\frac{{dE_{10}^{GUPI}}}{{dt}})}}} \right|_{Q = 0}} \simeq 9.97, {\left. {\frac{{(\frac{{dE_4^{GUPII}}}{{dt}})}}{{(\frac{{dE_{10}^{GUPII}}}{{dt}})}}} \right|_{Q = 0}} \simeq 9.97
\end{equation}
The results show evidently that the charged TeV-scale black holes radiate mainly into the $4$-dimensional brane independent of the type of the GUP. In fact, the charged TeV-scale black hole emits radiation both in the bulk and into the brane and the electric charge increases the radiation rate of the micro black hole into the brane.  We use the radius of the outer horizon to calculate the area of the black body emitter in equations (41) and (42). However, in the geometric optics approximation, the black hole acts as a perfect absorber of a slightly larger radius. Therefore, there is a critical radius ${r_c} = (\frac{{3\sqrt 3 }}{2}){r_0} \simeq 2.6{r_0}$ for a Schwarzschild black hole, in four dimensions for null geodesics, where ${r_0}$ is the event horizon radius. Detailed calculation have shown \cite{sanchez} that the total energy radiated is better approximated by assuming the area given by ${r_c}$ rather than ${r_0}$. Based on this argument, we draw attention that some corrections to equations (39) and (40) should be considered as equation (37) is related to the area. 
\\In this case, critical radius of the black hole as an absorber is given by \cite{emparan01} 
\begin{equation} \label{43}
{r_c} = {(\frac{{d - 1}}{2})^{\frac{1}{{d - 3}}}}\sqrt {\frac{{d - 1}}{{d - 3}}} {r_{h + }}.
\end{equation}
Based on the assumptions of the theory of large extra dimensions,  gravitations and possibly scalar fields, are the only types of the fields allowed to be emitted in the bulk during the Hawking evaporation phase. The emission on the brane can take the form of the fermions, gauge bosons, and scalar Higgs from the perspective of the brane observer. In this case, the radiation into the brane may lead to the experimental detection of the Hawking radiation and thus of the production of the TeV-scale black hole remnants. Finally, in the next section, we discuss the remnants and some charge effects on the micro black hole thermodynamics. 
\section{Black Holes Remnant and charge effect on thermodynamics}												
Since Zwicky's observation of the large velocity dispersion of the members of the coma galaxy \cite{zwicky}, the dark matter problem has been raised as part of the astrophysics for more than eighty years. However, it took several decades to be recognized as a real problem. Its modern form goes back to the early 1980's when the so-called cold dark matter paradigm appeared \cite{peebles}. Most of the matter in the universe is made up of the dark matter in which its identity is an open problem. So far, it has been observed only through its gravitational interactions. A logical possibility is that dark matter is hidden, that is, neutral under all standard model gauge interactions \cite{berezhiani,mohapatra,berezhiani01,foot}. There exist many dark matter candidates. In order to describe the properties of elementary particle candidates all the possible models use the standard concept of quantum field theory \cite{kikuchi,kusenko}. Most of the candidates are non-baryonic weakly interacting massive particles (WIMPs) or WIMP-like particles \cite{cavahli01}. Black Hole remnants are a natural candidate for dark matter \cite{mcgibbon} since they are a form of WIMPs \cite{cline}. Recently, it is investigated in which non-charged black hole remnants are the primary source of the dark matter based on the GUP model \cite{chen,santiago,mehdipour}. 
On the other hand, some models have been proposed in which dark matter particles are charged under some hidden gauge group \cite{ackerman,feng,hooper,kim,huh}. Here, based on the equations (25), and (28), we have shown, the final stage of evaporation of the charged TeV-scale black hole is a remnant which has mass increasing with spacetime dimensions.The allowed particles forming the black hole at LHC are quarks, antiquarks, and gluons which formed nine possible electric charge states such as $ \pm \frac{4}{3}$, $ \pm 1$, $ \pm \frac{2}{3}$, $ \pm \frac{1}{3}$, 0 \cite{gingrich,landsberg}. Figure \ref{fig3} shows the electric charge effects on the minimum mass of the black hole and the maximum Hawking temperature. In this case, when the electric charge increases, the minimum mass and its order of magnitude increases, and the temperature peak shifts to the lower temperature. This consequence allows us to divide the remnant into several categories. 
In case of $Q = 0$, the stable remnant of the micro black hole could be considered as WIMPs or WIMP-like particles \cite{chen,santiago,mehdipour}.
The ratio of the black hole charge, $Q$, with respect to its mass, $M$, generally could be considered as three states; i) $Q > M$ ii) $Q \le M$ iii) $Q \simeq M$. It can be shown that it is impossible to make the electric charge of a classical black hole larger than its mass, in Planck unit, by an influx of the charged particles on the horizon \cite{misner,lightman} and this result remains true in TeV-scale as well \cite{dolgov}. The charged black hole strongly prefer to emit particles of the same sign since they penetrate the potential barrier easier \cite{dai}. This allows the black hole to discharge its electric charge easily. In this manner, the electric charge goes to zero much faster than its mass and it has been shown \cite{dai} that for the dark matter particles whose the ratio of the charge to the mass is much less than one, it could exclude the heavy dark matter and the existence of primordial black hole is incompatible with the ratio of the charge to the mass which is of order one. It was calculated, that the charge and the mass of the charged particles which could be considered as a candidate of the dark matter, fall in the following range
\begin{equation} \label{44}
100\left ( Q/e \right )^{2}\leq M\leq 10^{8}\left ( Q/e \right )TeV
\end{equation}
then, their absence in the galactic disk can be naturally explained by their interaction with the galactic magnetic fields \cite{chuzhoy}. Therefore, evidently, the charged Tev-scale black hole remnant could be considered as the potential candidate of the dark matter. In this case, it is straightforward to calculate the limited range of the minimal length and maximal momentum coefficient, $\alpha$, based on the equation (25) and (44). 
\begin{table}[ph] 
	\tbl{Range of the minimal length and maximal momentum coefficient, $\alpha$, limited by the possible mass of the dark matter candidate for different values of the charge, dimensions $d = 4,5,6$ in GUPI.}
	{\begin{tabular}{@{}cccc@{}} \toprule
			$|Q|$ & $d = 4$ & $d = 5$ &
			$d = 6$ \\
			 \colrule
			1/3 & $0.13788 \le \alpha  \le 1.458 \times 10^8$ & $0.48727 \le \alpha  \le 1.164 \times 10^4$ & $0.77370 \le \alpha  \le 5.502 \times 10^2$ \cr
			\hline
			2/3 & $0.13759 \le \alpha \le 2.916 \times 10^8$ & $0.48630 \le \alpha  \le 1.646 \times 10^4$ & $0.77207 \le \alpha \le 6.933 \times 10^2$ \cr
			\hline
			1 & $0.13753 \le \alpha \le 4.374 \times 10^8$ & $0.48612 \le \alpha \le 2.015 \times 10^4$ & $0.77178 \le \alpha \le 7.936 \times 10^2$ \cr
			\hline
			4/3 & $0.13751 \le \alpha  \le 5.832 \times 10^8$ & $0.48606 \le \alpha  \le 2.327 \times 10^4$ & $0.77167 \le \alpha \le 8.735 \times 10^2$ \cr 
			\botrule
		\end{tabular} \label{ta1}}
\end{table}
Table 1 shows the range of the minimal length and the maximal momentum, $\alpha$, for different values of the charge in different dimensions, $d = 4,5,6$. It is obvious, that increase of the spacetime dimension  applies narrow cuts on the range of the $\alpha$ in presence of the GUPI. In this way, one can also calculate the range of the minimal momentum coefficient, $\beta$, based on equations (28) and (44) as well (see Table 2).     
\begin{table}[tph] 
	\tbl{Range of the minimal length and maximal momentum coefficient, $\alpha$, as well as minimal momentum coefficient, $\beta$, limited by the possible mass of the dark matter candidate for different values of the charge, dimensions $d = 4,5,6$ in GUPII.}
	{\begin{tabular}{@{}cccc@{}} \toprule
			$|Q|$ & $d = 4$ & $d = 5$ &
			$d = 6$ \\
			\colrule
			1/3 & ${0.21865 \le \alpha  \le 4.443 \times {{10}^7}}$ & $0.09485 < \alpha  \le 3.546 \times {{10}^3}$ & $0.23577 < \alpha  \le 1.676 \times {10}^2$ \cr
			\hline
			2/3 & $0.21703 \le \alpha \le 8.886 \times 10^7$ & $0.09473 \le \alpha \le 5.014 \times 10^3$ & $0.23528 \le \alpha  \le 2.112 \times 10^2$ \cr
			\hline
			1 & $0.21674 \le \alpha \le 1.333 \times 10^7$ & $0.094718 \le \alpha \le 6.141 \times 10^3$ & $0.23519 \le \alpha \le 2.418 \times 10^2$ \cr
			\hline
			4/3 & $0.21663 \le \alpha  \le 1.777 \times 10^7$ & $0.094710 \le \alpha  \le 7.092 \times {10}^3$ & $0.23515 \le \alpha  \le 2.661 \times {10}^2$ \cr
			\hline
			1/3 & $0.649 \times 10^ {-8} \le \beta  \le 1.32021$ & $0.814 \times {{10}^{ - 4}} \le \beta  \le 3.04333$ & $0.172 \times {{10}^{ - 2}} \le \beta  \le 1.22436$ \cr
			\hline
			2/3 & $0.324 \times {10}^{ - 8} \le \beta  \le 1.33008$ & $0.575 \times {{10}^{ - 4}} \le \beta  \le 3.04704$ & $0.136 \times {{10}^{ - 2}} \le \beta  \le 1.22693$ \cr
			\hline
			1 & $0.216 \times {{10}^{ - 8}} \le \beta  \le 1.33189$ & $0.470 \times {{10}^{ - 4}} \le \beta  \le 3.04772$ & $0.119 \times {{10}^{ - 2}} \le \beta  \le 1.22740$ \cr
			\hline
			4/3 & $0.162 \times {{10}^{ - 8}} \le \beta  \le 1.33252$ & $0.407 \times {{10}^{ - 4}} \le \beta  \le 3.04796$ & $0.108 \times {{10}^{ - 2}} \le \beta  \le 1.22757$ \cr 
			\botrule
		\end{tabular} \label{ta2}}
\end{table}
However, to calculate the range of the $\alpha$ and $\beta$, one needs to take into account ${\beta ^2} \le \frac{1}{{12{\alpha ^2}}}$ as another constraint (see equation (28)) rather than equation (44). In this case, it is obvious, that the range of the $\alpha$ has more limited in presence of the $\beta$ (Table 1 and 2). 
As a potential candidate for the dark matter, and based on the pioneering work of Chuzhoy et al \cite{chuzhoy}, one can consider the charged TeV-scale black hole remnant into three categories. The first one is the micro black hole remnant with positive unit charge which its chemical properties would be very similar to the proton and could recombined with the electrons. The second group is the TeV-scale remnant with negative unit charge which can recombine with the baryons, forming neutral or positively charged particles. The third category is the remnant with fractional charge which might also recombine with the ordinary matter, though with smaller binding energies in which the combinations would be more vulnerable to dissociation. Therefore, if the fundamental Plank scale is of the order of TeV, LHC would produce charged black hole which as a consequence of their evaporation yield the charged black hole remnants as a candidate of dark matter particles.
\section{Conclusion}
In this paper, through two different forms of the generalized uncertainty principle as our primary inputs, we have calculated the Hawking temperature, the Bekenstein-Hawking entropy of the charged TeV-scale black hole in the framework of the extra dimension scenario based on the ADD model. In scenario with the large extra dimensions applying GUP type I and II, the black hole temperature increases (Figure \ref{fig1} and \ref{fig2}). It is evident, the electric charge affects on the maximum Hawking temperature and the minimum mass of the black hole and also in the large extra dimensions scenario, the final stage of the evaporation has the mass more than its four dimensional counterpart, and yields the charged black hole remnants which could be considered as a potential candidate of the dark matter particles which applies more cutoffs and limitations on the $\alpha$ and $\beta$ (see Table 1 and 2). It is shown that when the electric charge increases, the minimum mass and its order of magnitude increase and the temperature peak shifts to the lower temperature (see Figure \ref{fig3}). We were able to calculate the radiation rate of the charged TeV-scale black hole. It was found that, the charged black hole radiate mainly into the brane in accordance with the results of the previous pioneering work \cite{emparan01}. It was found that the electric charge increases the radiation rate of the black hole in the brane. This feature leads the charged micro black holes to less classical behaviors. The consequence of the calculations is a more reasonable framework of the signatures of the TeV-scale quantum gravity. 

\section*{Aknowledgement}
We would like to thank the anonymous referees for these helpful comments. We think revising the paper according to their suggestions will be quite straightforward.The paper is supported by University of Malaya (Grant No. Ru-023-2014) and Universiti Kebangsaan Malaysia (Grant No. FRGS/2/2013/ST02/UKM/02/2).


\end{document}